\documentclass[12pt]{article}
\input epsf
\begin{document}

\begin{center}
{\Large \bf STRANGELETS IN TERRESTRIAL ATMOSPHERE}

\vspace{0.5in}

{\bf Shibaji Banerjee$^a${\footnote{email: phys@boseinst.ernet.in}},
 Sanjay K. Ghosh$^a${\footnote{email: phys@boseinst.ernet.in}},
Sibaji Raha$^a${\footnote{email: sibaji@boseinst.ernet.in}} and
Debapriyo Syam$^b$}

\vspace{0.5in}
(a) Department of Physics, Bose Institute, \\ 93/1, A. P. C. Road, Calcutta
700 009, INDIA \\
\vspace{0.2in}
(b) Department of Physics, Presidency College, \\
86/1, College Street, Calcutta 700 073, INDIA\\

\end{center}
\begin{abstract}
A new dynamical model for the propagation of strangelets through the
terrestrial atmosphere is proposed.
\end{abstract}

Strangelets are small lumps of strange quark matter (SQM) which consist of
roughly equal numbers of up, down and strange quarks. Since the suggestion of
Witten\cite{wit}in 1984 that these strangelets, and not ordinary nuclear
matter, represent the true ground state of QCD, this area of research has
expanded considerably. The existence of stable or metastable lumps of SQM
would have numerous implications for physics and astrophysics. Most
importantly, they can account for the cosmological dark matter problem to a
large extent, if not entirely \cite{ars}.
Moreover, the mere existence of strangelets may be one of the most perfect
evidences for the QGP phase transition that is believed to have occurred in
the early universe or inside neutron stars.
\par
To establish the existence of strangelets, it is, of course, necessary to
detect them. The obvious place to search for them would be in the cosmic
rays. To this end, it is necessary to understand the propagation of
strangelets through the earth's atmosphere, as these would have to traverse
the whole atmospheric depth before arriving at the detectors. In this letter,
we attempt to provide a realistic dynamical model which describes the
propagation of strangelets.
\par
The fact that SQM is absolutely stable does not contradict the ordinary
experience that matter is, for the most part, made of ordinary nuclear matter.
This is because, in order for ordinary matter to change to SQM, a large
number of $u$ and $d$ quarks need to change to $s$ quarks, which would
require a very high order weak interaction and therefore can be considered
to be highly improbable \cite{wit}. On the other hand, the existence of
ordinary nuclear matter shows that quark matter consisting of only $u$ and
$d$ quarks is unstable \cite{wit}. Introduction of a third flavour reduces
the energy relative to a two-flavour system since another Fermi well is now
present and this makes the system stable. In the earlier works \cite{fgm},
calculations along the lines of Fermi gas model seemed to suggest that
although strangelets with high atomic numbers would be absolutely stable,
the same could not be said for strangelets with low A; the surface effect
which must be taken into account for small strangelets would make the energy
per baryon increase with decreasing baryon number, turning the SQM unstable.
In later calculations \cite{shm1,shm2}, the Fermi gas model was replaced by
the spherical MIT hadronic bag models and the stability of strangelets with
low A was examined in the light of this model. The works done by various
authors \cite{an1,an2} tell us that for low $s$-quark masses, shell-like
structures occur for A = 6, 18, 24, 42, 54, 60, 84, 102 etc.These values may
change somewhat with the change in the strangeness fraction \cite{an1,an2}.
\par
Strangelets are expected to possess a small positive electric charge. They
would have been neutral, if the ground state composition consisted of equal
numbers of quarks of the three types of quark flavours, which is the most
favourable state. Actually, however, the $s$ quark is heavier than the other
two flavours and so their number is slightly less than that of the $u$ or
$d$ quarks. Hence the fortuitous cancellation
$\frac{2}{3}n_{u}-\frac{1}{3}n_{d}-\frac{1}{3}n_{s}$ does not occur exactly,
as a result of which we are left with a small residual positive charge.
It can also be argued from mere experience that they cannot be negatively
charged because although overall charge neutrality can be maintained by
covering them up by a cloud of positrons, they will eat up every piece of
ordinary matter they find in their path since it is energetically favorable
for nuclear matter to convert to SQM. (For a recent survey of the physics of
SQM, see \cite{mad}.)
On the other hand, a small positive charge would be helpful, as in that case,
ordinary nuclei will be electrostatically repelled when the strangelets are
moving slowly; when they move fast enough (highly energetic), nothing can
prevent them from absorbing neutrons and becoming more and more tightly bound.
\par
It is generally assumed that the strangelets which come to the upper layer
of the atmosphere have baryon number $A\sim 1000$ or more. This assumption is
usually made on the basis of the following experimental and theoretical
considerations. Some of the reports which suggest  direct candidates for SQM
are tabulated below.\\

\vspace{0.2in}
\begin{tabular}{|l|c|c|}    \hline
\em Event & \em Mass &  \em Charge  \\ \hline
Counter experiment \cite{tab1} & $A\sim 350-450$ &14  \\
Exotic Track \cite{tab2}& $A\sim 460$ & 20 \\
Price's Event \cite{tab3} & $A > 1000$ & 46 \\
Balloon Experiments \cite{tab41,tab42} & $A\sim 370$ & 14\\ \hline
\end{tabular}
\par
\vspace{0.2in}
The theoretical tip is usually taken from the generally accepted mass number
for strangelets which is given in the above paragraph. It is often remarked
that that the observation of these candidates at such large atmospheric depths ($\sim 500$ $g/cm^2$)
requires unusual penetrability of these lumps which means that their cross
sections should be very small and hence the geometric size much smaller than
typical nuclear size. This, however, is not borne out by the models of SQM
where the density of SQM is seen to be not much larger than ordinary nuclear
matter \cite{wit}.
\par
Wilk \textit{et al.} \cite{wlk1,wol,wlk2,wlk3} proposed a mechanism by which
the strangelets are able to cover great atmospheric depths. They assumed that the mass and hence
the cross sections of the strangelets incident on the upper layers of the
atmosphere with initial masses of the order of $10^3$ a.m.u. decrease rapidly
due to their collisions with air molecules in which a mass equal to that of
the nucleus of an atmospheric atom is ripped off from the strangelet in every
encounter. They also proposed that there should exist a critical mass
$m_{crit}$ such that when the mass of the strangelet evolving out of an
initially large strangelet  drops below that, it simply evaporates
into neutrons and that is what would fix the lower limit of the altitude
upto which a strangelet would be able to penetrate.
\par
Let us recapitulate the basic conclusions that can be derived from the
earlier works. Firstly, strangelets observed at the mountain
altitudes typically have mass around 300 to 450 and charge between 10 to 20.
But the experimental results obtained till date are inconclusive and hence
they do not impose a strict bound on the mass and charge of strangelets that
can be observed in future experiments. Secondly, although the correlation
between penetrability and geometric cross sections is usually valid for
ordinary nuclei, the same cannot be easily extrapolated to the case of
strangelets since these are very tightly bound massive objects and are not
expected to break up as a result of nuclear collisions. Indeed, in a typical
interaction between a strangelet and the nucleus of an atmospheric atom, it
is more probable for the strangelet to absorb neutrons so that the
colliding nucleus, and not the strangelet, is likely to break up most of the
time. Hence the scheme proposed in \cite{wlk1}, namely that the mass of a
strangelet decreases in every encounter, seems to be unrealistic. We, in
this letter, propose an alternative scheme based on the following premises :
\begin{enumerate}
\item The collision of a lump of SQM with ordinary matter results in the
\emph{absorption} of the neutrons from the colliding nucleus, as a result of
which the mass of the strangelet increases in every collision and it becomes
more tightly bound.
\item The initial masses of the strangelets are assumed to be small in
order to obtain final baryon numbers which are nearly equal to the observed
ones at mountain altitudes. The discussion above indicates that it is quite
possible to have stable lumps of SQM with low mass numbers.
This would also facilitate a somewhat larger flux in the cosmic rays.
\item The speed, and hence the kinetic energy of these particles, must be
such that they would arrive at a distance of 25 km above the sea level,
surmounting the geomagnetic effects. We start with such an altitude since 
the atmospheric density above 25 km is low enough to be neglected. The charge
of the strangelet is also fixed by this assumption, corresponding to a
certain strangeness fraction.

\end{enumerate}
The simple assumptions proposed above give a picture more or less in accord
with the observation of the propagation of the strangelets in the
atmosphere, which can give useful indications of the type of things to be
expected in an actual experiment. The description of the model is given next.
We consider a situation in which a strangelet with a low baryon number
enters the upper layers ($\sim 25 Km$ from the sea level) of the atmosphere.
To arrive at this point, a charged particle must possess a speed determined
by the formula (see, \textit{e.g}, \cite{menzel}).
\begin{equation}
\frac {pc}{Ze} \geq \frac{M}{r_{o}^{2}} \frac{cos^{4}\vartheta}{\left
( \sqrt{1+cos^{3}\vartheta}+1\right )^2}
\end{equation}
where $M$is the magnetic dipole moment of the Earth, $r_o$ the radius of
the Earth and $\vartheta$ is the (geomagnetic) latitude of the point of
observation ($\sim 30^o$, which might represent a location in north eastern
India). $p$ and $Ze$ represent the momentum and charge, respectively, of the
particle. The magnetic field of the Earth is taken to be equivalent to that
due to a magnetic dipole of moment $M=8.1\times 10^{22} J/T$, located near the
centre of the earth, the dipole axis pointing south. We have fixed the mass,
initial speed and charge to be 64 amu $6.6\times10^{7}$m per sec and 2
(electron charge), respectively, at the initial altitude of 25 km, for the
purpose of the present work.
\par
In the course of its journey, the strangelet comes in contact with
air molecules, mainly $N_2$. During such collisions, the strangelet absorbs
neutrons from some of these molecules, as a result of which it becomes more
massive. The effect of such encounters is summarized in the formula

\begin{equation}
\frac{dm_S}{dh}= \frac{f\times m_{N}}{\lambda}
\end{equation}
where $m_S$ is the mass of the strangelet, $m_{N}$ the total mass of the
neutrons in the atmospheric atom, $\lambda$  the mean free path of the
strangelet in the atmosphere and $h$ the path length traversed. (It should be
emphasized here that the strangelets would preferentially absorb neutrons,
as protons would be coulomb repelled. Nonetheless, some protons could still
be absorbed initially when the relative velocity between the strangelet and
the air molecule is large, leading to increase in both the mass and the
charge of the strangelet. We do not address this issue in the present work
for the sake of simplicity, although it can be readily seen that the rate of
increase in mass would obviously be faster than that in charge.) In the
above equation, $\lambda$ depends both on $h$, which determines the density
of air molecules and the instantaneous mass of the strangelet, which relates
to the interaction cross section. The mean free path  decreases as lower
altitudes are reached since the atmosphere becomes more dense and the
collision frequency increases. Finally the factor $f$ determines the fraction
of neutrons that are actually absorbed out of the total number of neutrons in
the colliding nucleus. The expression for this factor has been determined by
geometric considerations \cite{gosset} and is given by
\begin{equation}
f=\frac{3}{4}{\left(1-\nu\right)}^{1/2}{\left(\frac{1-\mu}{\nu}\right)}^{2}
-\frac{1}{3}{\left[3{\left(1-\nu\right)}^{1/2}-1\right]}
{\left(\frac{1-\mu}{\nu}\right)}^{3}
\end{equation}
In eqn(3), $\mu=\frac{b}{R_1+R_2}$ and $\nu=\frac{R_1}{R_1+R_2}$ where
$b$ is the impact parameter. $R_1$ and $R_2$ are the radii of the strangelet
and the nucleus of the atmospheric atom, respectively.
$f$ is initially small but grows larger and reaches the limiting value 1
when the strangelet grows more massive.
\par
The above considerations lead us to a set of differential equations of the
form
\begin{equation}
\frac{d\vec{v}}{dt}=-\vec{g}+\frac{q}{m_S}\left ( \vec{v} \times \vec{B}
\right ) - \frac{\vec{v}}{m_S}\frac{dm_S}{dt}
\end{equation}
In eqn(4), $-\vec{g}$ represents the acceleration due to gravity, $\vec{B}$
is the terrestrial magnetic field, $q=Ze$ and $\vec{v}$ represents the
velocity of the strangelet. These equations were solved by the $4^{th}$
order Runge Kutta Method for the set of initial conditions described above.
\par
The results are shown in figs.~ 1 -- 5. Figure 1 shows the variation of
Altitude with time, the zero of time being at 25 km. The time required to
reach a place which is about 3.6 km above the sea level (height of
a typical north east Indian peak like `Sandakphu', where an experiment to
detect strangelets in cosmic rays using a large detector array is being set
up \cite{saha}) is indicated in the figure.
The next figure (Fig.~2) shows the variation of the mass of the strangelet
with time and figure 3 shows the variation of the strangelet mass with
altitude. It can be seen from the figures that the expected mass at the
aforementioned altitude comes out to be about 340 amu or so. Figure 4 shows
the variation of the mean free time with altitude. Finally, figure 5 shows
the variation of $\beta=v/c$ with time, showing the expected decrease of the
speed with time, which also justifies the adequacy of the nonrelativistic
treatment that we have applied.

\begin{figure}
\begin{center}
\epsfbox{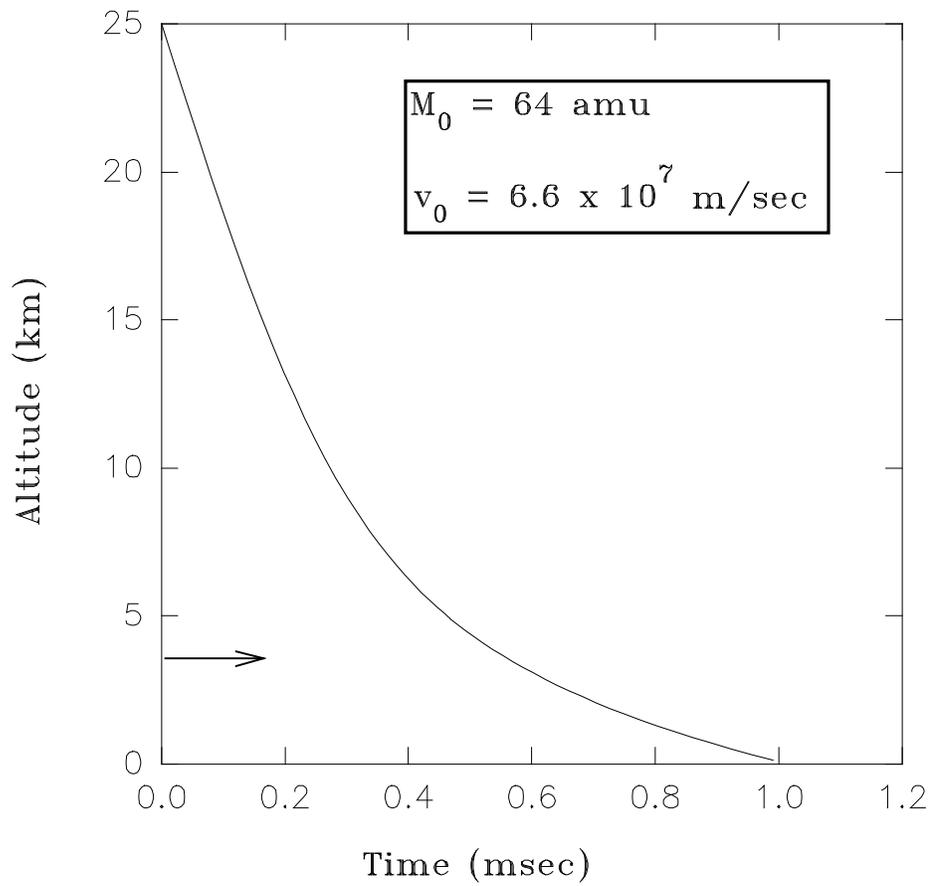}
\end{center}
\caption{Variation of altitude with time. The arrow corresponds to an
altitude of 3.6 km from the sea level.}
\end{figure}
\begin{figure}
\begin{center}
\epsfbox{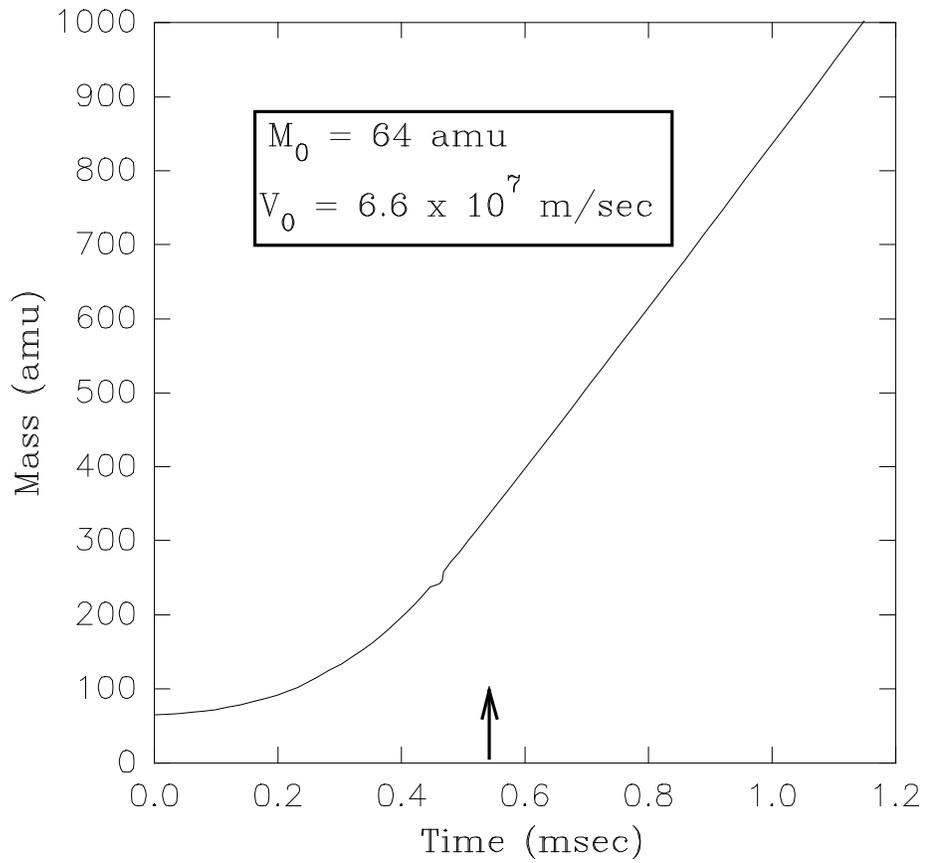}
\end{center}
\caption{Variation of mass with time. The time indicated by the arrow is the
time taken to reach an altitude 3.6 km from the sea level, starting from
25 km.}
\end{figure}
\begin{figure}
\begin{center}
\epsfbox{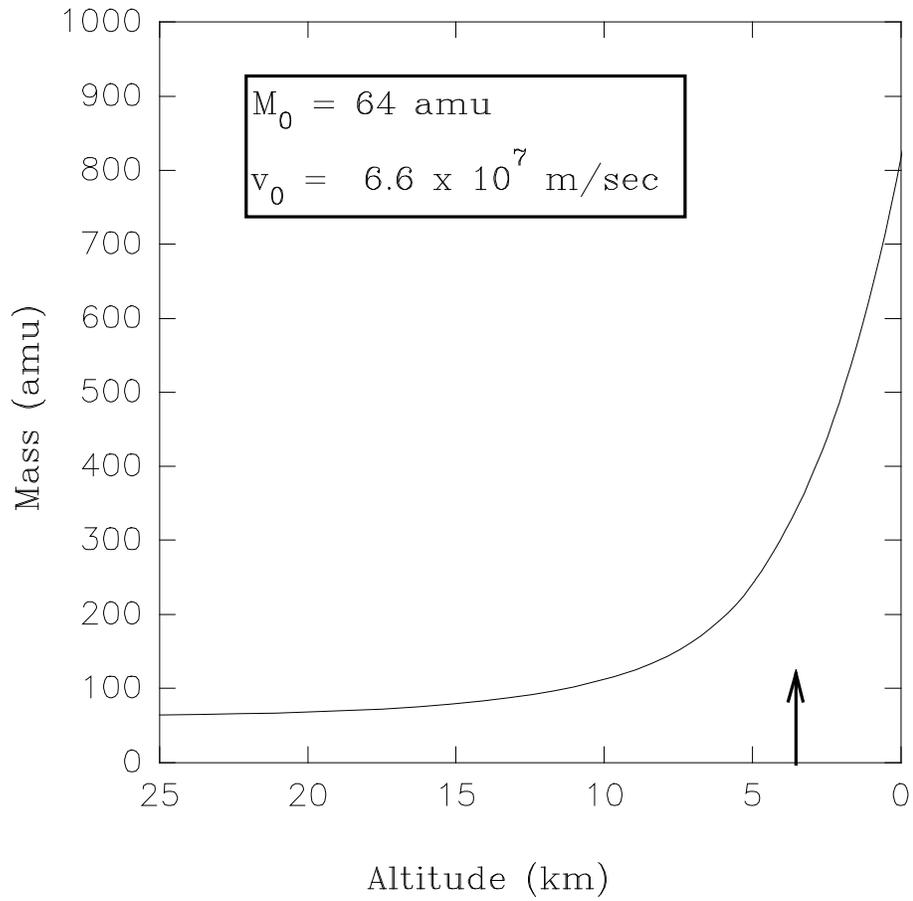}
\end{center}
\caption{Variation of mass with altitude. The arrow corresponds to  an
altitude of 3.6 km from the sea level.}
\end{figure}
\begin{figure}
\begin{center}
\epsfbox{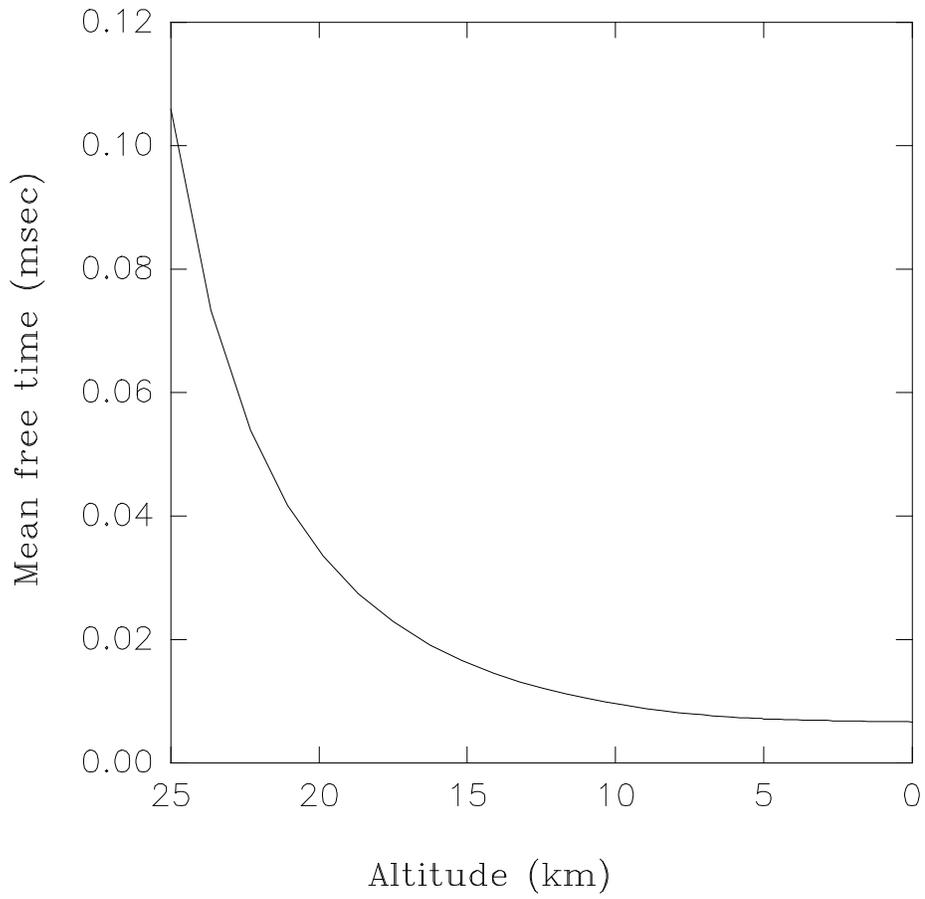}
\end{center}
\caption{Change of mean free time with altitude.}
\end{figure}
\begin{figure}
\begin{center}
\epsfbox{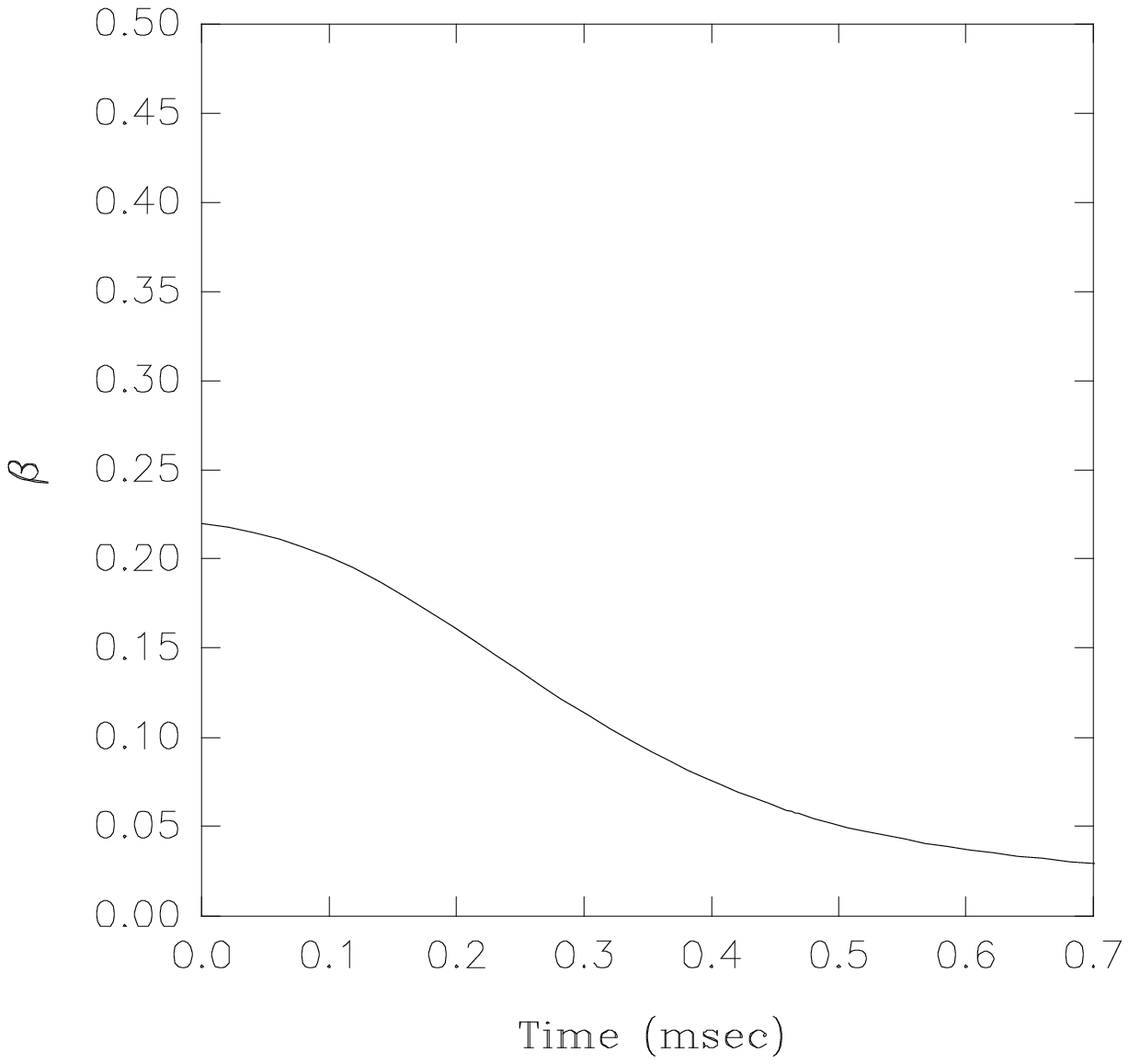}
\end{center}
\caption{Change of $\beta=v/c$ with time.}
\end{figure}

\par
In this letter, we have proposed a dynamical model of the propagation
of strangelets through the atmosphere of the Earth. Although the model is
based on simple assumptions, it is realistic enough to include the
difference in the interaction process which is expected when SQM, and not
ordinary nuclei, collide with the atmospheric nuclei. The effects of Earth's
gravitational and magnetic fields are also included in the equations of
motion so that meaningful information can be extracted directly from the
resulting trajectory. The main conclusion of the model is that the exotic
cosmic ray events with very small $Z/A$ ratios at mountain altitudes could
result from SQM droplets (strangelets) which need not be too large initially.
Thus the flux of strangelets in the cosmic rays may indeed be appreciable
enough to make their detection by a large area detector at mountain
altitudes a real possibility.
\par
The works of SB and SKG were supported in part by the Council of Scientific
\& Industrial Research, Govt. of India, New Delhi.



\begin{thebibliography}{99}
\bibitem{wit}E. Witten, \textit{Phys. Rev.} \textbf{D30}, 272 (1984)
\bibitem{ars}J. Alam, S. Raha and B. Sinha, \textit{Ap. J.} (in press)
\bibitem{fgm}E. Farhi and R. L. Jaffe, \textit{Phys. Rev.} \textbf{D30},
2379 (1984)
\bibitem{shm1}E. Farhi and R. L. Jaffe, \textit{Phys. Rev.} \textbf{D30},
1307 (1986)
\bibitem{shm2}E. P. Gilson and R. L. Jaffe, \textit{Phys. Rev. Lett}
\textbf{71}, 332 (1993)
\bibitem{an1}M. G. Mustafa and A. Ansari, \textit{Phys. Rev.} \textbf{D53},
5136 (1996)
\bibitem{an2}M. G. Mustafa and A. Ansari, \textit{Phys.Rev.} \textbf{C55},
2005 (1995)
\bibitem{mad}Jes Madsen,\textit{astro-ph} \textbf{9809032}; to appear in
\textbf{Hadrons in Dense Matter and Hadrosynthesis}, \textit{ Lecture Notes
in Physics}, Springer Verlag, Heidelberg.
\bibitem{tab1}M. Kasuya \textit{et al}., \textit{Phys. Rev.} \textbf{D47},
2153 (1993)
\bibitem{tab2}M. Ichimura \textit{et al}., \textit{Nuovo Cim.}
\textbf{A106}, 843 (1993)
\bibitem{tab3}T. Saito, \textit{Proc.$24^{th}$ ICRC Rome} \textbf{1}, 898
(1995)
\bibitem{tab41}O. Miyamura, \textit{Proc.$24^{th}$ ICRC Rome} \textbf{1},
890 (1995)
\bibitem{tab42}J. N. Capdeville \textit{et al}., \textit{Proc.$24^{th}$ ICRC
Rome} \textbf{1}, 910 (1995)
\bibitem{wlk1}G. Wilk and Z. Wlodarczyk, \textit{J. Phys.} \textbf{G22},
L105 (1996)
\bibitem{wol}E. Gadysz-Dziadus and Z.Wlodarczyk, \textit{J. Phys.}
\textbf{G23}, 2057 (1996)
\bibitem{wlk2}G. Wilk and Z. Wlodarczyk, \textit{Nucl. Phys. (Proc. Suppl.)}
\textbf{52B}, 215 (1997)
\bibitem{wlk3}G. Wilk and Z. Wlodarczyk, \textit{hep-ph} \textbf{9606401}
\bibitem{menzel} \textbf{Fundamental Formulas of Physics, Vol. 2},
\textit{Donald H. Menzel (Ed)}, p. 560, Dover Publications, New York (1960)
\bibitem{gosset} J. Gosset \textit{et al}., \textit{Phys. Rev.}
\textbf{C16}, 629 (1977)
\bibitem{saha} \textbf{Science at High Altitudes}, \textit{S. Raha, P. K.
Ray and B. Sinha (Eds.)}, Allied Publishers, New Delhi (1998)
\end{thebibliography}
\end{document}